\documentstyle[aps,preprint,prl,psfig]{revtex}
\tightenlines
\begin{document}
% \draft command makes pacs numbers print
\draft
% repeat the \author\address pair as needed
\title{Power-law Localization in 2D, 3D
with Off-diagonal Disorder}
\author{Shi-Jie Xiong$^1$ and S. N. Evangelou$^2$}
\address{$^1$ Department of Physics, Nanjing University,
Nanjing 210093, China \\
$^2$ Department of Physics, University of Ioannina, Ioannina
45110, Greece } \maketitle
\date{\today}
\begin{abstract}
% insert abstract here
We describe non-conventional localization of the midband 
$E=0$ state in square and cubic finite bipartite lattices 
with off-diagonal disorder by solving numerically 
the linear equations for the corresponding amplitudes.
This state is shown to display  multifractal fluctuations, 
having many sparse peaks, and by  scaling the participation ratio
we obtain its disorder-dependent fractal dimension $D_{2}$.
A logarithmic average correlation function 
grows as $g(r) \sim \eta \ln r$ at distance $r$ 
from the maximum amplitude and is consistent with a typical overall 
power-law  decay $|\psi(r)| \sim r^{-\eta }$
where $\eta $ is proportional to the strength 
of off-diagonal disorder.

\end{abstract}
% insert suggested PACS numbers in braces on next line
% body of paper here

%\section{INTRODUCTION}
It is well-known that off-diagonal disorder in the nearest neighbors 
of one-dimensional (1D) lattice is responsible for anomalous
localization of the corresponding state at the band center $E=0$
\cite{1,2,3,4}. The log-amplitude $\ln |\psi_i|$ at site $i$ 
executes a random-walk and the typical wave
function decays exponentially at distance $r$ from its maximum as
$|\psi(r)|\sim \exp(-\alpha \sqrt{r})$, with $\alpha$ a
disorder-dependent parameter. The $E=0$ is a special state arising from 
a sublattice symmetry, which can exist in the presence of off-diagonal
disorder and causes the Hamiltonian to change sign 
under the transformation $\psi_{i}\to (-1)^{i} \psi_{i}$. 
This symmetry exists in bipartite lattices with two sublattices, one
connected to the other \cite{1,5}, and is also
known as chiral symmetry. The corresponding
eigenvalues appear in pairs with energies $(E,-E)$ and since
the band center $E=0$ is a special energy the corresponding wave
function can be easily constructed.

\medskip
In 2D and 3D bipartite systems, such as square or cubic
lattices, the question of localization at the band center with
off-diagonal disorder is not resolved as in 1D and the decay of
the wave function amplitude from its maximum is not yet firmly
established. It is believed \cite{5,6} that
non-exponential decay occurs in 2D and even absence of decay in 3D
where extended states exist for weak disorder. 
The approach of \cite{5} predicted $\exp (-\gamma \sqrt{\ln r})$ decay
in every dimension, which is rather weaker than a power-law.
The theoretical interest in this problem
has been recently revived. On one hand, due  to the
related problem of random Dirac fermions \cite{7,8} which,
however, starts from a different zero disorder limit. In the case of
off-diagonal disorder one has a Fermi surface with many
points forming a square while in the problem of Dirac fermions
without disorder at zero energy four points exist only.
In the former case the pure density of states at $E=0$ 
has the well-known two-dimensional $\log$-singularity 
while in the Dirac case the density of states approaches zero
at the band center \cite{8}. On the other hand, the study of
localization with off-diagonal disorder could be
important for understanding properties of several realistic 2D
systems, such as states in semiconductor quantum wells and high
mobility silicon MOSFETs \cite{9}, current-currying states in
quantum Hall effect \cite{10}, etc.

\medskip
In this Letter we present an exclusive study of the midband $E=0$
state in 2D (squared) and 3D (cubic) bipartite lattices with
disorder in the nearest-neighbor hoppings. We compute the
coefficients of the wave function for large finite systems and obtain
the participation ratio. 
The  amplitudes are shown to display multifractal fluctuations 
with many scattered peaks, characterized by a fractal dimension
which depends on the strength of disorder. 
An appropriate correlation function is shown to behave 
as $g(r) \sim \eta \ln r$ where $\eta $ also depends on the strength of
disorder. It implies a  power-law decay $|\psi(r)| \sim r^{-\eta
}$ of the peak heights from the maximum peak, in contrast to the
one-dimensional typical decay $|\psi(r)|\sim \exp (-\alpha \sqrt{r})$.

\medskip
%\section{THE  COMPUTATION OF THE $E=0$ STATE}
We consider a tight-binding Hamiltonian in  2D, 3D
lattices with random nearest-neighbor hoppings
\begin{equation}
H=\sum_{\langle ij \rangle }(t_{ij}c^{\dag }_ic_j+\text{H.c.}),
\end{equation}
where $c_i$ is the annihilation operator of an electron on the
$i$th site and the nearest-neighbor hopping integrals $t_{ij}$
are real random variables satisfying the probability distribution
\begin{eqnarray}
P(\ln t_{ij}) = \frac{1}{W} \, \, \, \, \,
\text{  for  }-\frac{W}{2} \leq \ln
t_{ij} \leq  \frac{W}{2}.
\end{eqnarray}
This disorder, commonly used in 1D, guarantees positive hoppings 
and the typical variation of the random elements $W$ 
is believed to be a good measure for the strength 
of off-diagonal disorder in any dimension \cite{11,12}. 
If, instead, we choose to distribute directly $t_{ij}$ 
between $-W/2$ to $+W/2$ the disorder can never be made 
strong enough since there is no energy scale 
in the Hamiltonian. In the following we consider 
off-diagonal disorder form Eq. (2) and ignore 
diagonal disorder by setting the site energies equal to zero.

\medskip
A bipartite lattice consists of two interconnected 
sublattices $A$ and $B$ with the random hoppings $t_{ij}$ 
connecting the sites of one sublattice to the sites 
of the other. It was shown \cite{1} that a finite system
with $n_A$ sites on sublattice $A$ and $n_B$ sites on 
sublattice $B$ ($n_B>n_A$) has at least $n_B-n_A$ linearly
independent eigenfunctions with eigenvalues exactly $E=0$.
Moreover, the amplitudes of these states vanish on the sites of 
sublattice $A$.  In the present study we consider finite square 
and cubic lattices with their $L$-site edges 
along the main directions in 2D and 3D.
In the calculations $L$ is odd and Dirichlet
boundary conditions are applied with sublattice $A$ having
$\frac{L^2-1}{2}$ ($\frac{L^3-1}{2}$) sites and sublattice
$B$ has $\frac{L^2+1}{2}$ ($\frac{L^3+1}{2}$) sites for the
square (cube). According to Ref. \cite{1} only one 
$E=0$ state exists in this case with  finite amplitude 
on sublattice $B$ and zero amplitude on sublattice $A$
and the corresponding amplitudes satisfy the linear
system of equations 
\begin{equation}
\label{linear}
 \sum_{\delta ,(j+\delta \in B)}t_{j,j+\delta }
\psi_{j+\delta }=0, \text{  for all } j\in A,
\end{equation}
where $\delta $ is summed over the nearest neighbors of site $j$
so that $j+\delta$ belongs to the sublattice $B$ where the amplitude
is finite. The total number of equations 
$\frac{L^2-1}{2}$ ($\frac{L^3-1}{2}$) derived from the $A$ sites
allows to obtain the $\frac{L^2+1}{2}$ ($\frac{L^3+1}{2}$) amplitudes on
the $B$ sites. This procedure suffices to determine uniquely
the  $E=0$ state since the the normalization 
condition $\sum_i |\psi_i|^2=1$ accounts for the one missing equation.

 \centerline{\psfig{figure=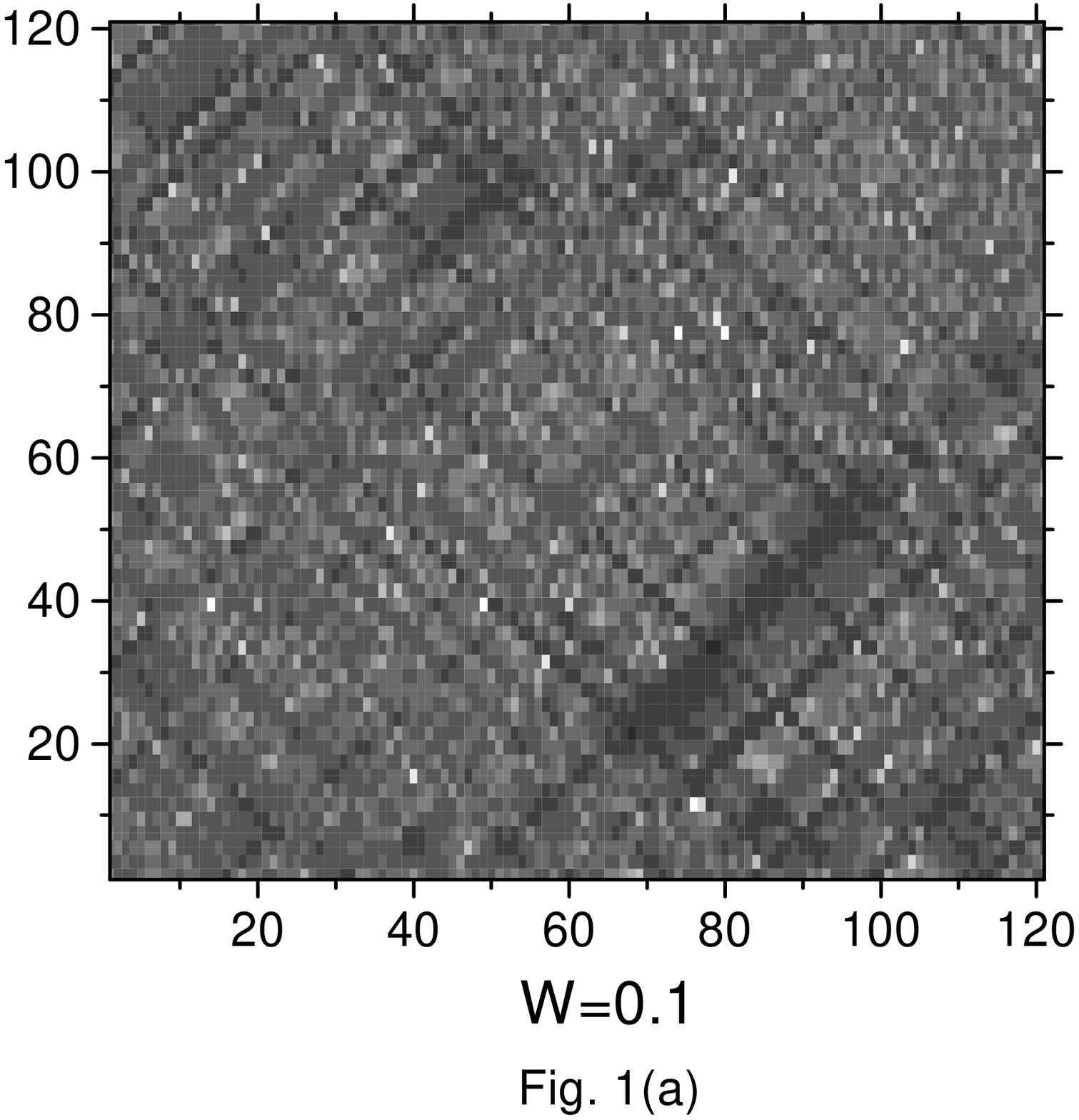,width=3.in,height=4.3in}}
 \centerline{\psfig{figure=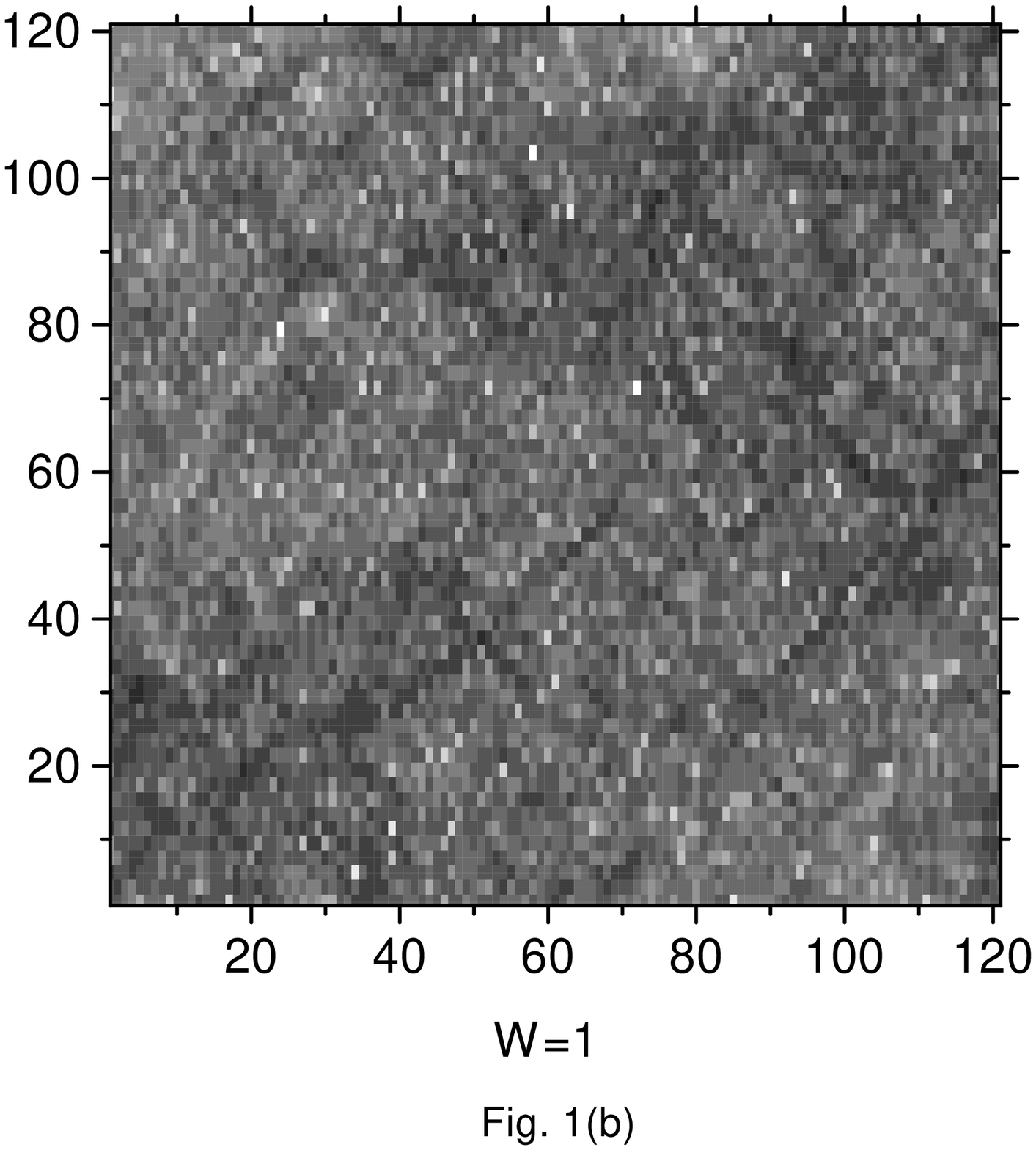,width=3.in,height=4.3in}}
 \centerline{\psfig{figure=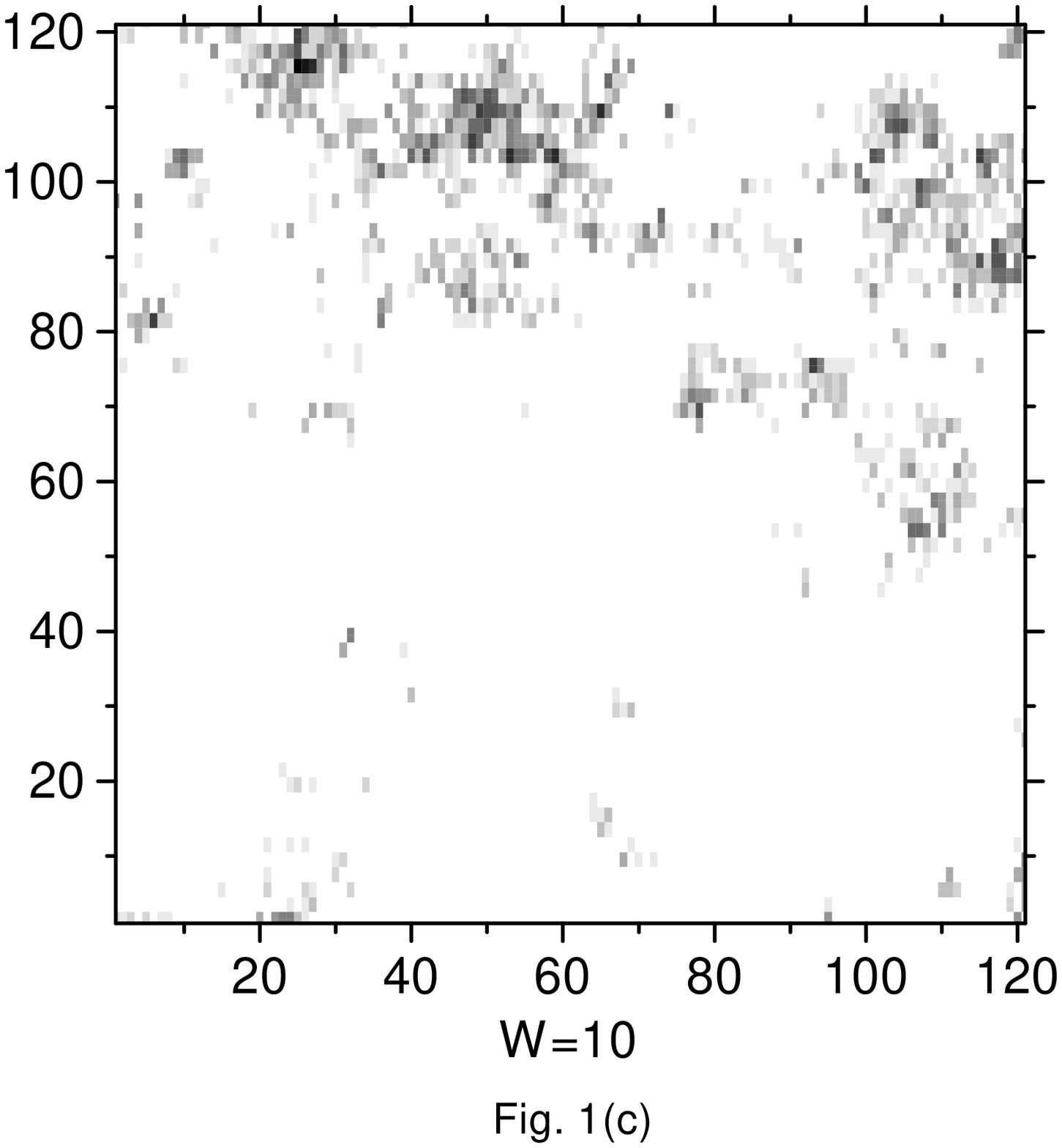,width=3.in,height=4.3in}}
{\footnotesize{{{\bf Fig. 1.} The logarithmic amplitude $\ln
|\psi|$ of the $E=0$ wave function in 2D squared lattices 
with off-diagonal disorder strength (a) $W=0.1$, (b) $W=1$ and (c) $W=10$. 
The darkest regions cover areas with the higher $\ln |\psi|$, between $0$ and
$-1$, and the lightest regions areas with the lowest $\ln |\psi|$,
between $-12$ and $-13$, while the various degrees of gray denote 13 linear
scales.
 }}}
 \par

\medskip
%\section{THE FRACTAL CHARACTER OF THE $E=0$ STATE}
We have solved numerically the linear system of equations (3) 
for different strengths $W$ of the off-diagonal disorder. 
In Fig. 1 typical pictures of the logarithmic amplitudes 
of the $E=0$ wave function in 2D are shown which 
display fractal characteristics. Similar fractal
patterns are obtained for all examined disorder strengths 
in 2D and 3D. 
The logarithmic amplitude for very weak disorder (small $W$)
is periodically distributed in Fig. 1(a), 
the fractal character of the state is obvious 
for intermediate disorder $W$ in Fig. 1(b) 
and for higher values of disorder 
(large $W$) the area with significant amplitude 
becomes vanishingly small fraction of the total in Fig. 1(c). 
The latter case implies stronger decay of the $E=0$ state, similarly
to what one expects for ordinary localized states with 
strong diagonal disorder.
However, the important difference to conventional localization
is that the maximum is not concentrated in a small region of space,
but many maxima of almost similar heights exist. 
These peaks with relatively large amplitude are randomly scattered over
space and exist also for large disorder (Fig. 1(c)). Therefore,
the $E=0$ state in the presence 
of off-diagonal disorder shows unusual localization properties.

\medskip
Firstly, we investigate the fractal structure of the normalized 
wave function by computing the inverse participation ratio
\begin{equation}
p=\sum_i |\psi_i|^4 \sim L^{-D_2},
\end{equation}
where the scaling of $p$ with size $L$ defines the fractal
dimension $D_2$. It is well-known that extended states 
are space filling with $D_{2}=d$, the space dimension,
and localized states are point-like with $D_{2}=0$. 
The probability distributions of $\ln p$ for various finite-size $L$ systems 
over a sufficient number of disorder configurations are shown in Fig. 2 
to be similar to the ones reported in \cite{13}.  
For example, in Fig. 2(b) we plot the distributions for the intermediate
disorder strength $W=1$,  which are roughly
invariant in shape as the size $L$ gets larger and
a sudden drop occurs for small values of $\ln p$ 
with a tail for large $\ln p$.
We focused on the peaks of such distributions
by determining $p_{max}$ from the maximum probability of $p$ 
as suggested in \cite{13}. The $p_{max}$ 
shift towards lower values by increasing
the size so that the variation of this typical value
of $p$ according to Eq. (4) is shown in the insets of
Fig. 2 where we plot $\ln (p_{max})$ vs. $\ln L$. The 
fractal dimension for  $W=1$ is $D_2 \approx 1.55$. 

\medskip
The behavior of this special state
for off-diagonal disorder is, somehow,  reminiscent
of the conclusions reached in \cite{14} where the 
2D wave functions are multifractal in the presence 
of disorder for length scales shorter than
the localization length. For off-diagonal disorder 
a similar situation is encountered since the
usual (exponential) localization length 
is guaranteed to diverge at $E=0$.
In Fig. 2 we plot the probability distribution of the
inverse participation ratio for $W=0.1$, $W=1.0$ and $W=10$,
where $D_{2}$ is well-defined 
from the slope of the 
$\ln p_{max}$ vs.  $\ln L$ almost linear curve
in the investigated sizes.
For  small disorder $W=0.1$ we find
$D_{2}\approx 2$, which implies an almost extended state,
on the contrary, for strong disorder $W=10$
the logarithm of the typical inverse participation ratio is close
to zero, which implies strong localization with $D_{2}\approx0$.

 \centerline{\psfig{figure=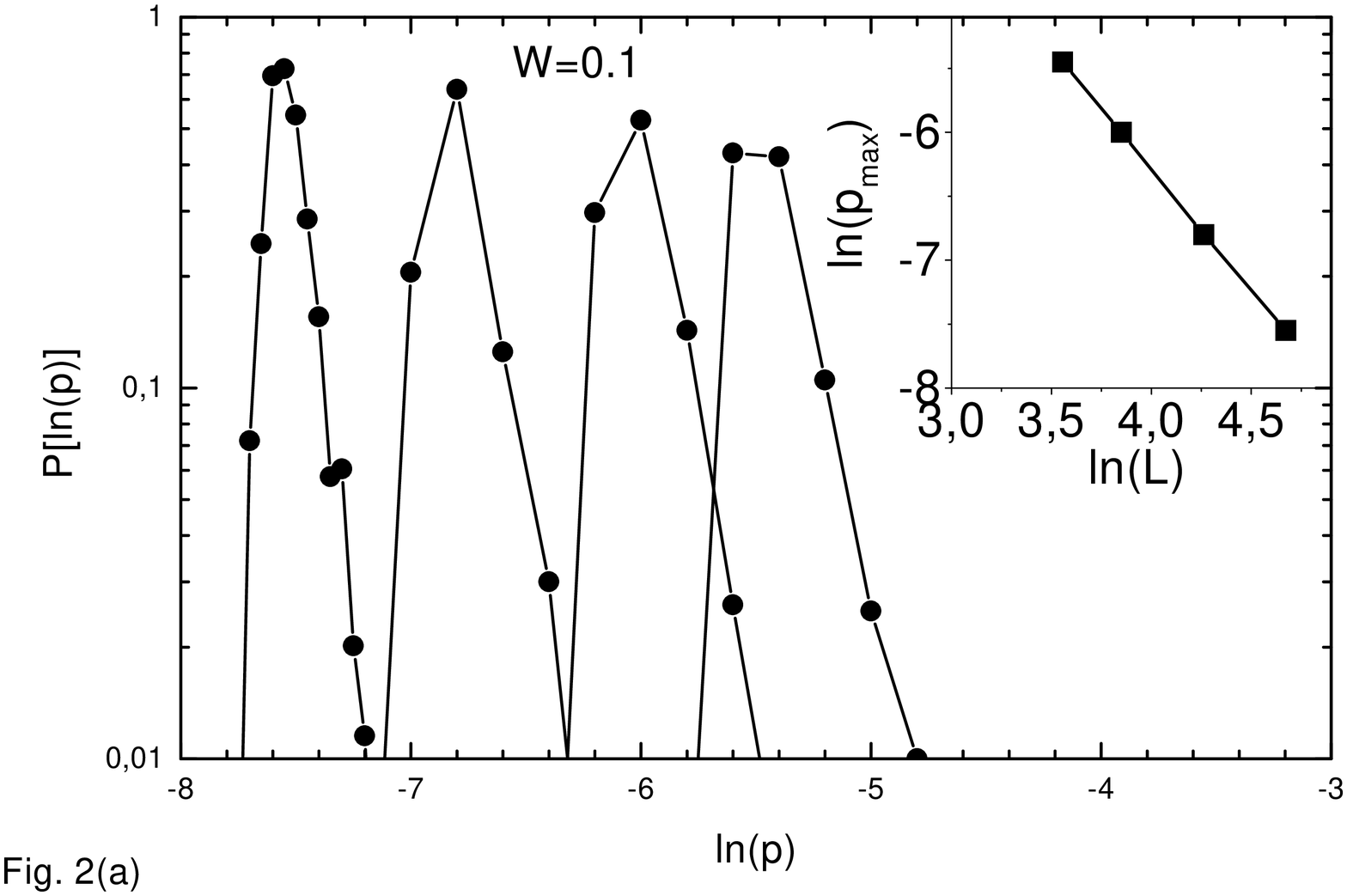,width=4.in,height=2.4in}}
 \centerline{\psfig{figure=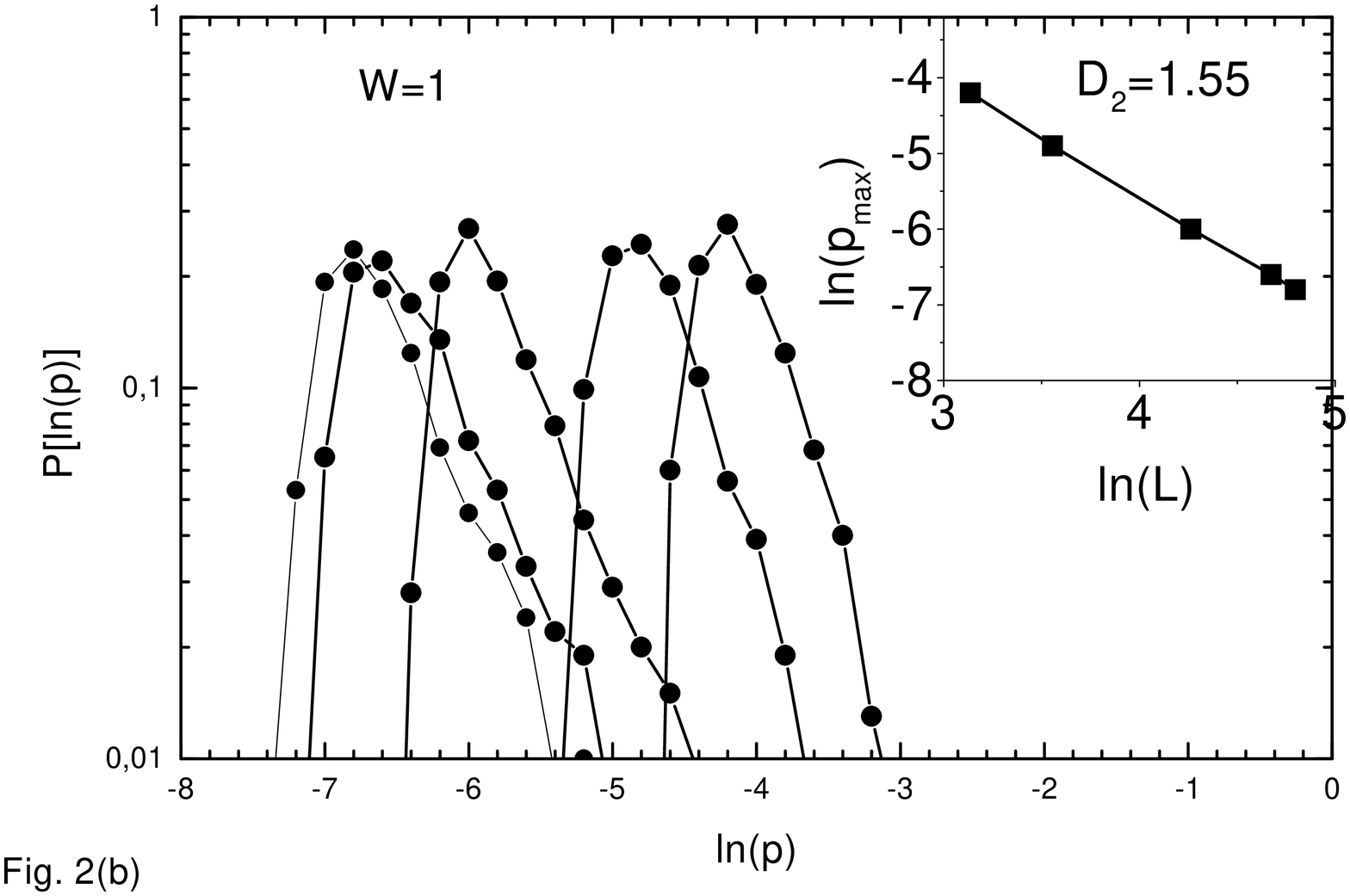,width=4.in,height=2.4in}}
 \centerline{\psfig{figure=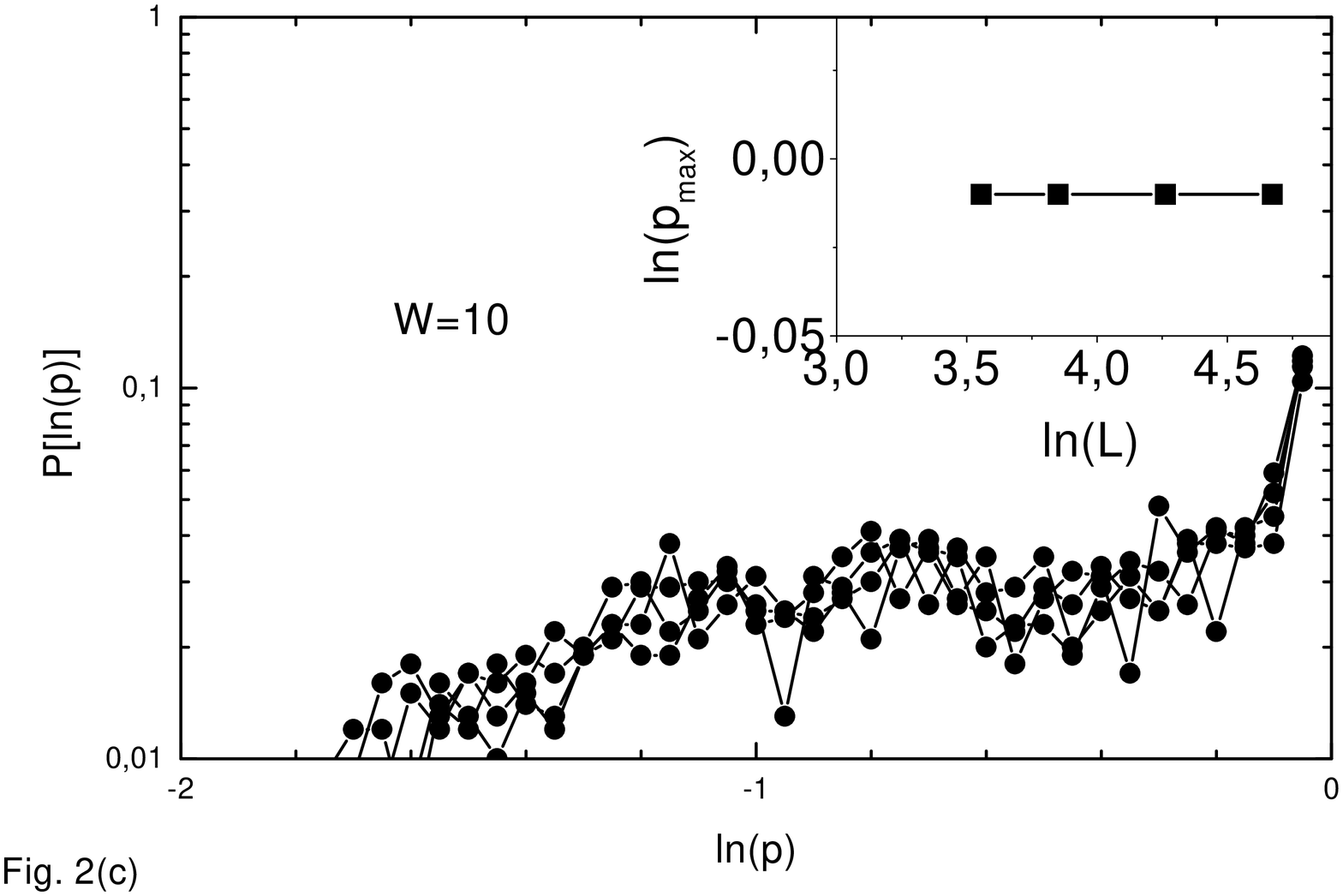,width=4.in,height=2.4in}}
{\footnotesize{{{\bf Fig. 2.} The probability distribution for
the logarithm of the inverse participation ratio for the $E=0$ wave
function in square (2D) lattices with sizes $L=23$; $35$;
$71$; $107$ and $121$.  The inset shows $\ln (p_{max})$ vs. $\ln L$
for (a) $W=0.1$, (b) $W=1$, (c) $W=10$.
 }}}

\medskip
%\section{POWER-LAW TYPICAL DECAY OF THE $E=0$ STATE}
Finally, we examine the typical decay of the $E=0$ state
by introducing an appropriate correlation function. 
Following the definition of \cite{1} we have studied 
\begin{equation}
g(r)=\langle \ln {\frac {|\psi_{i_{max}}|} {|\psi_j|}}
\rangle  \text{,  }
r=|{\bf r}_j-{\bf r}_{i_{max}}|,
\end{equation}
where $i_{max}$ is the site of the lattice with the largest
amplitude and the average $\langle \ldots \rangle $ is taken over
a sufficient number of random configurations 
and various spatial directions. In 
1D bipartite lattice, where the log-amplitude of the $E=0$ state 
with off-diagonal disorder exhibits random-walk behavior,
the correlation function is $g(r) \sim \sqrt{r}$ for large $r$
\cite{1}.  In Fig. 3 we plot $g(r)$ vs. $\ln r$ in 2D by increasing $W$.
These curves for not too small $r$ can be reasonably well fitted to
the relation $g(r)=\eta \ln r +\beta $, 
with parameters $\eta $ and $\beta$. 
For example, for $W=1$ we find $\eta =0.44 \pm 0.01$,
$\beta =1.51\pm 0.02$ and the average amplitude 
asymptotically decays from its maximum as a power
law $|\psi(r)|  \sim r^{-\eta }$.
We conclude that the
multifractal amplitude pattern shows many random
peaks and the power-law decay reflects
the long-range decay of the peak heights from the highest peak.
The amplitude for short distances from each peak
decays very sharply, instead.

  \centerline{\psfig{figure=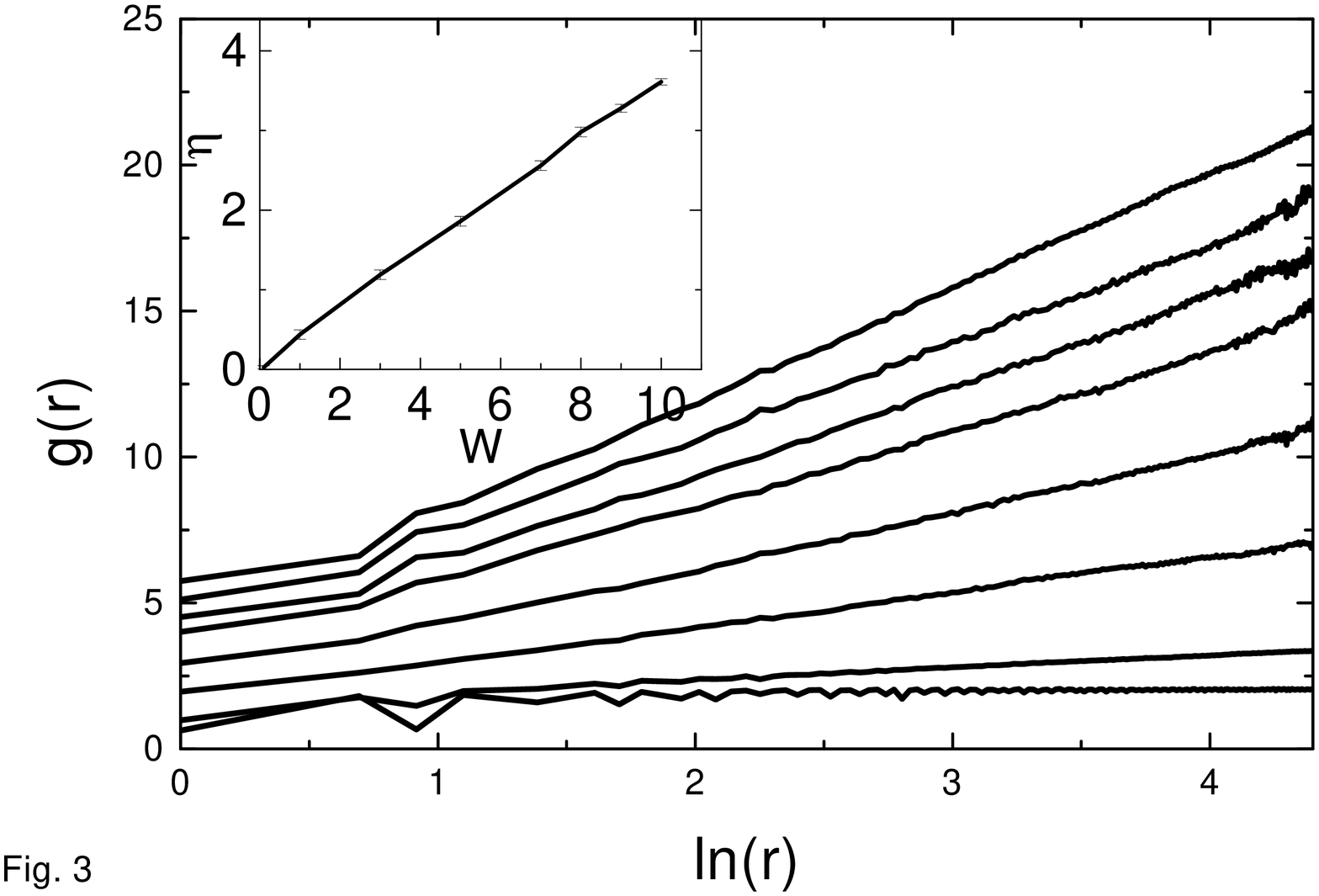,width=4.1in,height=2.8in}}
 {\footnotesize{{{\bf Fig. 3.} The correlation function $g(r)$
 versus $\ln r$ in 2D for $W=0.1,\, 1,\, 3,\, 5,\, 7,\, 8,\, 9,\,
 10$, from bottom to top and system-size $L=121$. All curves can be
 fitted by straight lines $g(r) \sim \eta \ln r+\beta $
 and the power-law exponent $\eta $ is plotted as a function of $W$ 
 in the inset.
 }}}
 \par

\medskip
We have also obtained similar relations $g(r)=\eta \ln r +\beta $
for the cubic (3D) case by plotting $g(r)$ vs. $\ln
(r)$ for different values of disorder where 
$g(r)$ still increases linearly with $\ln r$.
This means that the corresponding amplitudes
show a power-law decay from the maximum, for any finite $W$,
with the exponent $\eta $ proportional to $W$.
The unusual power-law localization of the mid-band state is rather
surprising in 3D, since nearby states 
might be extended if the disorder is weak.  
In fact, $\eta$ for both 2D, 3D,  is found to increase 
almost linearly with $W$, shown
in the insets of Fig. 3, 4, where we establish the very approximate 
relations $\eta \sim W/3$ in 2D and $\eta \sim W/9 $ in 3D.

  \centerline{\psfig{figure=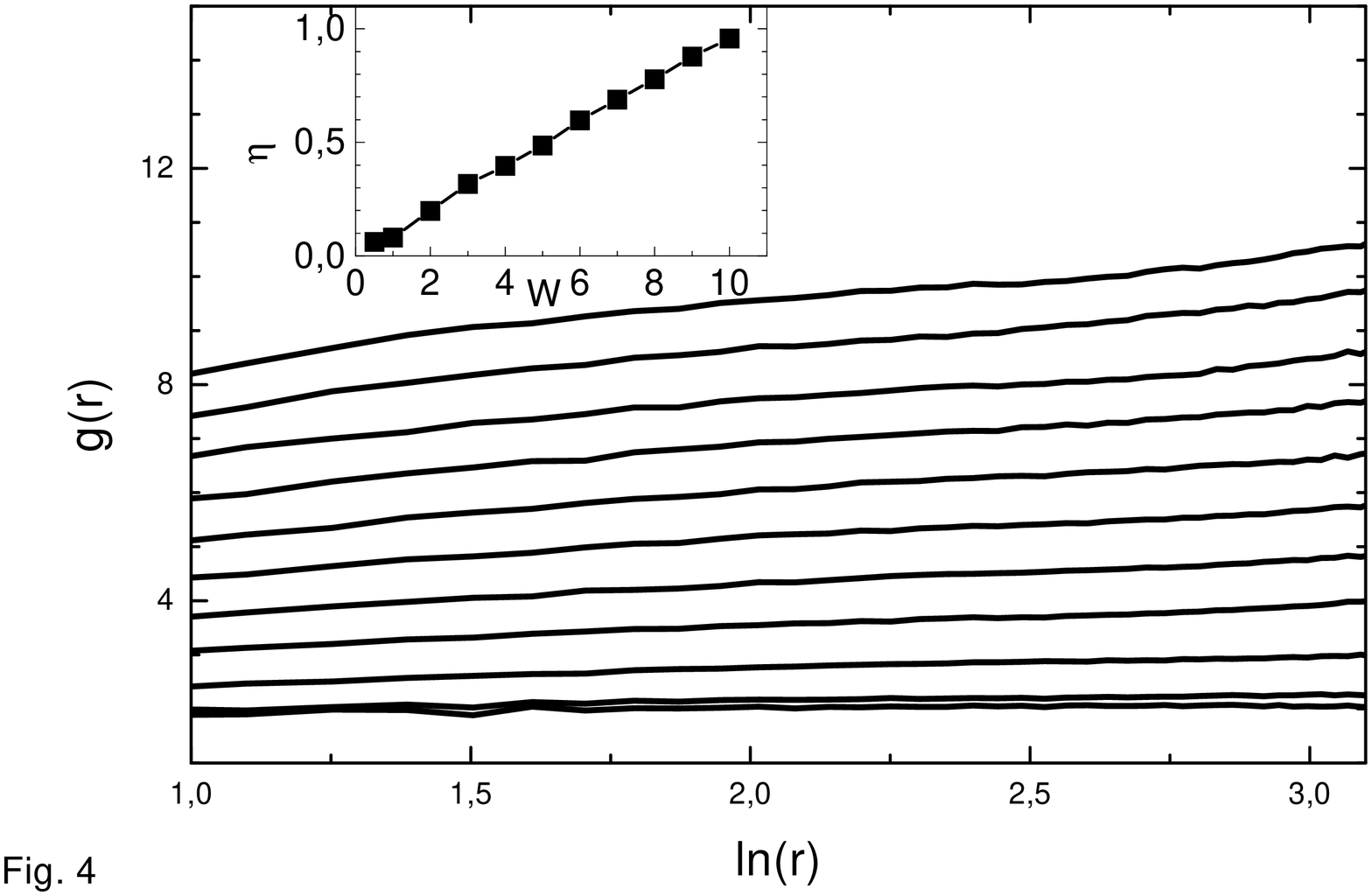,width=4.1in,height=4.in}}
 {\footnotesize{{{\bf Fig. 4.} The correlation function $g(r)$
 versus $\ln r$ for 3D with $W=0.1,\, 1,\, 2,\, 3,\, 4,\,
 5,\, 6,\, 7,\, 8,\, 9,\,10$, from bottom to top and
 system-size $L=21$. The curves can be fitted by straight lines $g(r) \sim
 \eta \ln r+\beta $ and the obtained $\eta $ is plotted in the inset 
 as a function of $W$.
 }}}
 \par

\medskip
%\section{DISCUSSION}
In summary, we have investigated the $E=0$ wave function for
square and cubic lattices with off-diagonal disorder. This 
zero energy wave function in the adopted geometry exists
for any disorder configuration and can be easily
studied numerically.  The wave function amplitude displays 
many sharp peaks randomly scattered in space and
is characterized by the fractal dimension $D_{2}$ 
which shows a strong dependence on the strength of disorder.
The amplitude from each peak falls off very rapidly 
for short ranges but the heights of the
peaks decay slowly from the main maximum via the 
power-law $r^{-\eta }$, where  $\eta$ is  
almost linearly proportional to the strength of disorder $W$.

% now the references. delete or change fake . delete next three
%   lines and directly read in your .bbl file if you use bibtex.

\end{document}